\RequirePackage[l2tabu, orthodox]{nag}
\documentclass[envcountsect,10pt]{llncs}
\usepackage[charter]{mathdesign}

\usepackage[T1]{fontenc}
\usepackage[utf8]{inputenc}
\usepackage{textcomp} 
\PassOptionsToPackage{hyphens}{url}
\usepackage[bookmarksnumbered,bookmarksopen,breaklinks,colorlinks=true]{hyperref}
\usepackage[hyphenbreaks]{breakurl}
\usepackage{graphicx}
\usepackage{overpic}
\usepackage{amsmath}
\usepackage[english]{babel}
\usepackage[style=numeric,maxnames=5,giveninits=true,doi=false,isbn=false,url=false,eventdate=comp]{biblatex}
\usepackage[short]{bibstrings}
\usepackage{xspace}
\usepackage{verbatim}
\usepackage{general}
\usepackage{markup}
\usepackage{annot}
\usepackage{bodyversion}
\usepackage{environment}
\usepackage{hyphenation}

\AtEveryBibitem{%
\ifentrytype{inproceedings}{
  \clearname{editor}%
  \clearfield{booktitleaddon}%
}{}
}

\renewbibmacro{in:}{%
  \ifentrytype{article}{}{\printtext{\bibstring{in}\intitlepunct}}}
\DeclareFieldFormat[misc]{title}{\mkbibquote{#1}}
\DeclareSourcemap{
  \maps{
    \map{
      \step[typesource=tweet, typetarget=misc, final]
      \step[fieldset=howpublished, fieldvalue={Twitter}]
    }
    \map{
      \step[typesource=blog, typetarget=article, final]
      \step[fieldset=entrysubtype, fieldvalue={newspaper}]
      }
    \map{
      \step[typesource=movie, typetarget=misc, final]
      }
    \map{
      \step[typesource=newspaper, typetarget=article, final]
      \step[fieldset=entrysubtype, fieldvalue={newspaper}]
      }
    }
}
\newcommand{\src}[2]{\footnote{See \url{#1} (Accessed #2).}}
\newcommand{\func}[1]{{\mathsf{#1}}}
\newcommand{\tuple}[2]{\langle #1,#2 \rangle}
\newcommand{\errmargin}{\sigma}
\newcommand{\horizon}{\Delta}
\newcommand{\duration}{T}
\newcommand{\distance}{d}
\newcommand{\epoch}[1]{\epsilon_{#1}}
\newcommand{\epochrand}[1]{e_{#1}}
\newcommand{\hash}{h}
\newcommand{\hashdom}{H}
\renewcommand{\privkey}[1][{}]{{k_{#1}}}

\renewcommand{\pubkey}[1][{}]{{K_{#1}}}
\newcommand{\enc}[2]{\widetilde{E}_{\pubkey[#1]}(#2)}
\newcommand{\xenc}[2]{E_{\pubkey[#1]}(#2)}
\newcommand{\rrencpart}[3]{\xenc{#2}{#3 \| \hash(\pubkey[#1])}}
\newcommand{\rrenc}[4]{\enc{#1}{\rrencpart{#1}{#2}{#3} \| #4}}
\newcommand{\rrencsym}[4]{\reflectbox{E}_{#1;#2}(#3)(#4)}
\renewcommand{\sign}[2]{[#2]_{\privkey[#1]}}
\newcommand{\ident}[1][{}]{{\id{id}_{#1}}}

\newcommand{\logdb}[2][{}]{\mathcal{L}_{#1}[#2]}
\newcommand{\broadcasttag}[1]{\textbf{#1}}
\newcommand{\broadcast}[2]{\langle \broadcasttag{#1},#2 \rangle}
\newcommand{\contactnow}[3]{#1 \approx_{#3} #2}
\newcommand{\contact}[2]{\func{contact}(#1,#2)}
\newcommand{\contactlist}[1]{\func{contacts}_{#1}}
\let\mid=\|
\def\|{\,\mid\,}
\newcommand{\popetsfigure}{}

\title{Hansel and Gretel and the Virus\thanks{%
    Version: Sun Feb 21 22:37:48 2021 +0100 / arXiv2-4-gf558ca9 /
    hansel-and-gretel.tex}
}
\subtitle{Privacy Conscious Contact Tracing}
\author{Jaap-Henk Hoepman\inst{1,2}}
\institute{
  Radboud University Nijmegen, Email: \email{jhh@cs.ru.nl} \and
  University of Groningen
}
\begin{document}
\maketitle
\begin{abstract}
  Digital contact tracing has been proposed to support the health authorities in fighting the current Covid-19 pandemic. In this paper we propose two centralised protocols for digital contact tracing that, contrary to the common hypothesis that this is an inherent risk, do not allow (retroactive) tracking of the location of a device over time. The first protocol does not rely on synchronised clocks. The second protocol does not require a handshake between two devices, at the expense of relying on real-time communication with a central server.

  We stress that digital contact tracing is a form of technological solutionism that should be used with care, especially given the inherent mass surveillance nature of such systems. 
\end{abstract}
\section{Introduction}

The road to hell is paved with good intentions. This is perhaps especially the case in exceptional times like these, as the world is suffering from a global pandemic caused by the Severe Acute Respiratory Syndrome CoronaVirus 2 (SARS-CoV-2). The pandemic has spurred the development of \term{contract tracing} apps, that aim to support the health authorities in their quest to quickly determine who has been in close and sustained contact with a person infected by this virus~\autocite{who2020contact-tracing,martin2020demystifying-covid19-tracing}. 

Contact tracing (also known as \term{proximity tracing} or \term{exposure notification}) apps have been in use in China for some time~\autocite{mozur2020corona-china-app}. The main idea underlying digital contact tracing is that many people carry a smartphone most of the time, and that this smart phone could potentially be used to more or less automatically collect information about people someone has been in close contact with. The work of Feretti~\etal~\autocite{ferretti2020covid}, modelling the infectiousness of SARS-CoV-2, showed that digital contact tracing could in principle help reduce the spread of the virus, under a number of strong assumptions:
\begin{itemize}
\item
  a significant fraction of the population use the app,
\item
  the app has a very low false negative rate,
\item
  a large fraction of infected people are successfully isolated, and
\item
  the delay between establishing a contact, determining whether that person is indeed infected, and getting that person to (self)-isolate is short (1 day or less).
\end{itemize}
The often cited threshold of $60$\% of people that need to install the app for digital contact tracing to be successful~\autocite{hinch2020effective} seems to be based on the rather optimistic assumption that people go into isolation without delay ($0$ days), there is a false negative probability less than $0.1$, and that $70$\% of infected people successfully isolate (looking at the data from Feretti~\etal). This is not to say that some reduction in the spread of the virus can already be observed at lowed adoption rates~\autocite{howell2020sixty-percent}.

Notwithstanding these rather optimistic assumptions, and with testing capacity initially insufficient to even test people working in high risk environments (like health care professionals), Bluetooth-based contact tracing apps have quickly been embraced by governments across the globe as a necessary tool to ease or end the lock-down as imposed in many countries in 2020. Such Bluetooth based contact tracing apps broadcast an ephemeral identifier on the short range Bluetooth radio network at regular intervals, while at the same time collecting such identifiers transmitted by other smartphones in the vicinity. The signal strength is used as an estimate for the distance between the two smartphones, and when this distance is determined to be short (within $1$--$2$ meters) for a certain period of time (typically $10$--$20$ minutes) the smartphones register each others ephemeral identifier as a potential risky contact. Some, like the European Commissioner for Internal Market Thierry Breton, even went so far as to call them `deconfinement' apps\footnote{%
  In a tweet on April 22, 2020,  \url{https://twitter.com/ThierryBreton/status/1253039000782286848}.
}.


This is a dangerous form of `technological solutionism'~\autocite{morozov2013click-here}: instead of thinking the underlying problem through, we go for a quick technological fix to counter the symptoms. This is by no means a Luddite argument to forego the use of technology per se, but rather a warning that  ``what typically separates good from bad practice is adherence to rigorous, contextual testing and working within existing expertise, institutions and approaches to augment what we know works, rather than to work around or challenge the integrity of those experts or knowledge''~\autocite{mcdonald2020digital-response}. Technological support for contact tracing works much better when it is actually developed in close coordination with the health authorities~\autocite{redmiles2020fix-contact-tracing}.

The model of Feretti~\etal is based on a very incomplete picture of how the virus spreads exactly. Even worse, the product lead of the TraceTogether app used in Signapore, warns that no Bluetooth contact tracing system deployed or under development, anywhere in the world, is ready to \emph{replace} manual contact tracing~\autocite{bay2020no-panacea}. The problem being that to establish whether two people that were in close contact could indeed have infected one another depends on context, like whether the place they met was properly ventilated. Such information about a context in which people met cannot be derived or even approximated using Bluetooth (or other commonly available sensors). A human-in-the-loop is therefore necessary.

Privacy by design is another good intention that paves the road to this particular hell. Even though many have proposed privacy conscious designs for a contact tracing system (see section~\ref{sec-related}), such systems \emph{are} a mass surveillance tool by definition. Which makes the framing of such apps as deconfinement apps all the more worrisome, even cynical. Especially in cases where the system's primary purpose is to take care of our health and well-being, there is a tendency to lower the bar and accept greater level of (state) surveillance as a form of `pastoral power'~\autocite{foucault1981omnes,golder2007foucault-pastoral}. As noted by Taylor \etal~\autocite{taylor2020data-justice-covid19} ``the pandemic has amplified a nascent epidemiological turn in digital surveillance''. Attempts to use the pandemic as justification for extending surveillance powers should be resisted~\autocite{eyal2020trolley-zealots}.


Notwithstanding these reservations, this paper will discuss yet another privacy conscious design for a contact tracing app. (And because of these reservations I hesitate to write privacy friendly in this context.) Why study contact tracing at all, given these reservations, one might well ask. There are several answers to that question. First of all, and what got us started on the subject many months ago, was the purely engineering question of whether contact tracing could be done with some level of privacy consideration in the first place. Second of all, it was clear from the start that governments and health authorities would consider digital technologies to support their work in controlling the spread of the virus. Resisting the ill-advised, perhaps desperate, reach for a technological quick fix is only a first line of defence. If that defence fails, we had better finished a comprehensive map of the design space and have the least problematic solution ready for them to grab.

\subsection{Contributions}

A common distinction among contact tracing apps is whether \emph{all} contacts are registered \emph{centrally} (on the server of the national health authority for example) or in a \emph{decentralised} fashion (on the smartphones of the users that installed the contact tracing app)~\autocite{martin2020demystifying-covid19-tracing}.\footnote{%
  Note that essentially all systems for contact tracing require a central server to coordinate some of the tasks. The distinction between centralised and decentralised systems is therefore \emph{not} made based on whether such a central server exist, but based on where the matching of contacts takes place.
}
In the first case, the authorities have a complete and perhaps even real time view of the social graph of all participants. In the second case, information about one's contacts is only released (with consent) when someone tests positive for the virus.

Our first contribution is to show that there actually exists a third semi-centralised category, between centralised and decentralised that \emph{only} provide the health authorities with the identities of those people that have been in close contact with an infected person.\footnote{%
  The DESIRE protocol, discussed in section~\ref{sec-related} is similarly semi-centralised, and was developed in parallel (when the first draft of this paper was written in March 2020, see \url{https://blog.xot.nl/2020/03/25/hansel-and-gretel-and-the-virus-privacy-conscious-contact-tracing/index.html}).
}
We present two solutions to the contact tracing problem that belong to this category. The first protocol is peer-to-peer based, but requires an actual exchange of messages between two devices before a contact is registered. On the other hand this protocol does not allow other phones (or the authorities for that matter) to later retroactively track the movements of a device over time. This refutes the hypothesis (specifically, risk SR7 identified in~\autocite{dp3t-privsec}) that this is an inherent, systemic, risk of all centralised contact tracing systems. The second protocol does not require a handshake between two devices, at the expense of relying on real-time communication with a central server and requiring synchronised clocks to prevent relay attacks.

In the course of our construction we also discuss a new cryptographic primitive called \term{replay resistant encryption} that prevents replay of encrypted messages in anonymous settings where senders cannot prove freshness of messages with a signature over a nonce.

We study semi-centralised instead of decentralised solutions to the contact tracing problem for two reasons. First of all, the discussion of the strong assumptions under which contact tracing could help reduce the spread of the virus we offered above indicates that a centralised solution (where the health authorities obtain a full and immediate picture of all people that have been in contact with an infected person) is preferable over a decentralised approach (that relies on the initiative of people to report to the health authorities themselves once the system has notified them of having a risk of infection). (Semi-)centralised solutions also offer the option to speed up the contact tracing process~\autocite{white2021decentralised-fail}. Second of all, the perceived privacy benefit of decentralised solutions is debated~\autocite{vaudenay2020dp3t}.

The paper is structured as follows. We first describe the main source for inspiration of this research, namely Apple's `find-my-iPhone' feature, in section~\ref{sec-inspiration}. We then formalise the contact tracing problem in section~\ref{sec-problem}.
After that we present and analyse two protocols in section~\ref{sec-prot1} and section~\ref{sec-prot2}. Section~\ref{sec-related} discusses related work. We finish with our conclusions in section~\ref{sec-conclusion}.


\section{Source of inspiration}
\label{sec-inspiration}

Our solutions described below are inspired by solutions to a related problem, namely that of locating lost devices. In particular, we use ideas present in Apple's `find-my-iPhone' feature that allows lost devices without a GPS sensor (like certain iPads or iWatches) to be located as well. This system supposedly\footnote{%
  No official documentation about how this feature is implemented appears to exist.
}
works as follows\src{https://blog.cryptographyengineering.com/2019/06/05/how-does-apple-privately-find-your-offline-devices/}{2020-03-25}.

The system distinguishes four different entities. $L$ is the lost device. $P$ is the trusted partner device the lost device was previously paired with. This is typically a different device owned by $L$'s owner. People are assumed to immediately pair new devices with devices they already own: without this earlier pairing, lost devices cannot be found. $H$ represents any of the millions of helper devices out there that are used by the system to locate lost devices. Finally, $C$ is the central cloud storage used by the helpers to store the location of (lost) devices they came in contact with, and that is used by the trusted partner devices to retrieve the location where their lost devices were last seen.

Device $L$ has a secret identifier $\id{id}_L$ known to $P$. Paired device $P$ has a public key $K_P$ known to $L$. $L$ uses its secret identifier $\id{id}_L$ to generate a random pseudonym $\id{rid}_L$, for example by hashing the identifier and the current time $t$ using a cryptographic hash function $h$, \ie $\id{rid}_L=h(\id{id}_L,t)$. $L$ broadcasts this random pseudonym every once in a while over its local Bluetooth and/or WiFi interface, in the hope that some helper $H$ (e.g. anybody else's iPhone that happens to be nearby) receives this signal. Randomising the identifier is necessary to prevent anybody from singling out and tracking $L$ through its static identifier $\id{id}_L$. (This assumes the MAC address of the Bluetooth or WiFi interface is properly randomised as well, which is not always the case~\autocite{martin2017mac-addr-randomisation}.)

This helper $H$ is supposed to know its current location (because it does have a GPS receiver) and is supposed to send its location and the pseudonym of the device it discovered to Apple's iCloud servers. The trusted partner $P$ trying to locate $L$ will query these servers with the recently used pseudonyms for $L$ using its knowledge of $\id{id}_L$ that allows him to compute $\id{rid}_L=h(\id{id}_L,t)$ for any $t$ in the recent past. If a tuple for any of these queries exists, the locations contained in it will be reported back as a result, allowing $P$ to locate $L$.

In order to protect the location privacy of the helper $H$, the location it reports to Apple's iCloud needs to be encrypted (otherwise Apple would know the location of all helpers), and it needs to be encrypted against $P$'s public key so $P$ can actually decrypt it. The only way for $H$ to know this key, is if the lost device $L$ broadcasts it alongside its randomised pseudonym. But as $P$'s public key is static, this would serve as an ideal static identifier to track $L$, thwarting the use of the randomised pseudonym. This can only work if $L$ can also randomise $P$'s public key $K_P$ before broadcasting it. Luckily, encryption schemes exist that allow the public key to be randomised while ensuring that the same unchanged private key is the only key that can be used to decrypt a message~\autocite{verheul2017pep}.

It is worth noting that this protocol appears, at first sight, to be vulnerable to devices that \emph{pretend} to be lost but instead are planted at specific locations. Such planted devices could lure helpers to encrypt their location to a `fake' public key $K_P$ broadcast by them, which is in fact controlled by the authorities that have the corresponding private key $k_p$. Luckily, Apple never releases the identity of the helper to the paired device. In other words, the scheme works because it is entirely under Apple's control. We need to trust Apple anyway as Apple could, at any time, decide to harvest the location of all devices it manufactures by a simple operating system update. 

\section{Problem definition}
\label{sec-problem}

Contact tracing is one of the traditional tools used to fight an epidemic. The goal of contact tracing is to find all recent contacts of a patient that tested positive for infection by a virus. The idea being that these contacts could have been infected by this patient. By tracing these contacts, testing them, and treating or quarantining them, spread of the virus can be contained~\autocite{who2020contact-tracing}. Contact tracing is typically done `by hand' by the national health authorities, and is a time consuming process. Digital contact tracing is supposed to support this traditional form of contact tracing, as people may not necessarily know or remember all people they have recently been in contact with.

Informally speaking, a system for contact tracing should help the health authorities to find all people that have been in close contact to a person carrying an infectious disease (like COVID-19), with the understanding that having been in close contact for a certain period of time with such a patient creates a significant enough risk of being infected as well. A contact is considered `close' if two people were less that several meters away from each other (denoted $\distance$ further on) for a certain duration (denoted $\duration$ further on). Such a contact is only relevant if it occurred at most $\horizon$ days before a person got tested positive for infection (which includes a possible delay between being infected and actually getting tested positive).

As many (but certainly not all) people carry a (smart) mobile phone with them almost all the time, it is natural to consider the use of such devices to \emph{automatically} collect data for contact tracing\footnote{%
  A low tech, non automatic, approach would be to ask people to keep track of all other people they meet in a small (paper) diary, as initially suggested by Prime Minister Jacinda Ardern of New Zealand, see:~\url{https://www.theguardian.com/world/2020/apr/19/jacinda-ardern-asks-new-zealanders-to-keep-diaries-to-help-trace-coronavirus-contacts}.
}.
As the precision required to establish closeness is in the range of at most several meters, GPS or mobile phone location data cannot reliably be used for this. Currently considered solutions for automatic contact tracing therefore quickly converged to the use of Bluetooth, even though even that technology has serious limitations~\autocite{dehaye2020bluetooth}. For non-automatic contact tracing, QR-code based solutions are being proposed\footnote{%
  E.g. the NZ COVID Tracer app, see:~\url{https://www.health.govt.nz/our-work/diseases-and-conditions/covid-19-novel-coronavirus/covid-19-novel-coronavirus-resources-and-tools/nz-covid-tracer-app}, or Zerobase, see:~\url{zerobase.io} 
}
that require users to explicitly scan QR codes of buildings they enter, or on the phone of other people they meet.

\subsection{Assumptions}

We assume a system of \term{mobile devices}, that move around freely, and that are always active. In the following we will assume smartphones as the primary devices.

Each device $X$ has a unique \term{identifier} $\ident[X]$, only known to itself, that it obtains by enrolling in the system. This identifier is inextricably linked to the unique owner of the device (that may get infected, or may need to be tested for infection). We therefore identify an owner with the identifier of their\footnote{%
  We use them/their to avoid the clumsy ``he or she'' or ``his or her'' gender neutral expressions.
}
device. In particular we write `infected device' $X$ for a  device $X$ that belongs to an infected person. We assume the authorities can contact the owner when given this identifier. Identifiers are secret and cannot be forged or guessed.

Each device is equipped with a \term{short range broadcast (radio) network} with which it can detect all other devices in the vicinity and communicate with them. In the following we will assume a Bluetooth network. It is assumed that all devices within maximal `infection distance' $\distance$ can be reliably detected and reached, and that the distance can be reliably measured over this network. That is to say, a message $m$ broadcast over the radio network reaches all devices within $\distance$ of the broadcasting device. Devices receiving this message as the tuple $\tuple{m}{d}$, where $m$ is the message received, and $d$ is the distance between the recipient and the sender\footnote{%
  We wish to stress, again, that this commonly made assumption is not at all realistic in practice.
}.
We do not assume that receivers are able to identify the sender of a message, or that they are able to determine whether two messages received were sent by the same device (unless the contents of the messages themselves would provide that information). This models the case where Bluetooth networks use frequently changing random MAC addresses. Devices can accurately measure passage of time; we do not assume synchronised clocks for the first protocol.

We furthermore assume a \term{central server} (operated by the health authorities $A$) with a public key $\pubkey[A]$ and corresponding private key $\privkey[A]$. All devices know the public key $\pubkey[A]$ of the authorities. Devices can reliably communicate in real time with this server using a \term{long range data network}. In particular, the server is authenticated (but the devices are not for privacy reasons), and the communication is encrypted. In the following we assume an Internet connection based on a cellular data network or a WiFi connection.

\subsection{Cryptographic primitives, and replay resistance}

The protocols below rely on two different semantically secure public key encryption schemes, for example based on elliptic curve cryptography.

One of these schemes, denoted $\enc{}{m}$, is used to encrypted messages broadcast on the Bluetooth channel. For this a 'raw' ElGamal based system based on Curve25519~\autocite{bernstein2004curve25519} is an appropriate choice, offering the equivalent of $128$ bits `symmetric' security~\autocite{ecrypt2018keylength,RFC7748}. With this choice, keys are $256$ bits ($32$ bytes) long, and so is the message space. Encryption is done using ElGamal directly (and not in a hybrid fashion where ElGamal is used to encrypt a random symmetric key that is then used to encrypt the message). Note that this means messages first need to be encoded as a point on the curve, see \eg~\autocite{DBLP:conf/acisp/FouqueJT13}. Ciphertexts are still semantically secure under the Decisional Diffie-Hellman assumption~\autocite{DBLP:conf/pkc/TsiounisY98}, but note however that this traditional method of using ElGamal directly is malleable and not IND-CCA (\ie indistinguishable against chosen ciphertext attacks~\autocite{katz2015modern-crypto}). This is not a problem in our protocols. The other scheme, denoted $\xenc{}{m}$, is an arbitrary strong IND-CCA cipher.

A cryptographic hash function  $\hash : \hashdom \mapsto \hashdom$ is also used. SHA-3 could be used for this~\autocite{FIPS202}.

One particular concern in our protocols is to prevent message replay attacks that try to create false positives (see the ``authenticity'' requirement in section~\ref{ssec-secpriv}). A standard way to ensure freshness of messages is to let the sender sign a fresh nonce (either provided by the receiver, or derived from the current time). It is not clear how to do so in the setting studied here were devices are supposed to remain anonymous (unless they have been in contact with an infected device).

In the first protocol below, devices $R$ broadcast their public key $\pubkey[R]$ and expect to receive information from nearby devices encrypted against this key. Adversarial devices may broadcast their own key, receive information from nearby devices encrypted against this key, decrypt this information and rebroadcast it (now encrypted against the key of another unsuspecting device) in a relay attack attempt. The specific information sent by a device in our protocols is actually its identity encrypted against the public key of the authorities, \eg $\xenc{A}{\ident[B]}$. We make use of the involvement of the authorities to implement \term{replay resistant encryption}, a construct that may be of independent interest.

In replay resilient encryption we distinguish three entities: an anonymous sender $S$, a relay $R$ (with key pair $\privkey[R],\pubkey[R]$) and the authorities $A$ (with key pair $\privkey[A],\pubkey[A]$). The goal is to let $A$ accept a message $m$ sent by $S$, but only if it was indeed received by $R$ directly from $S$. We write $\rrencsym{R}{A}{m}{m'}$ (where $m'$ is a message that is received by $R$ in the process; $R$ does not learn $m$).

One way to implement this primitive is to define
\[
\rrencsym{R}{A}{m}{m'} = \rrenc{R}{A}{m}{m'}~.
\]
Relay $R$ decrypts this message to obtain $m'$ and $M = \xenc{A}{m \| \hash(\pubkey[R])}$. It forwards $M$ to $A$ for decryption, together with its public key $\pubkey[R]$. $A$ then uses his private key $\privkey[A]$ and the public key $\pubkey[R]$ just received to verify the decryption of $M$, rejecting it if this does not contain $\hash(\pubkey[R])$.

A malicious adversary $Z$ trying the attack outlined above would receive $\rrenc{Z}{A}{m}{m'}$, which it can decrypt to obtain
$\xenc{A}{m \| \hash(\pubkey[Z])}$ and $m'$. Trying to replay this message against another device $R$ means sending
$\enc{R}{\xenc{A}{m \| \hash(\pubkey[Z])} \| m'}$, which $R$ can indeed decrypt, but which would fail the later test performed by $A$ (as
$\hash(\pubkey[Z]) \neq \hash(\pubkey[R])$). This prevents $Z$ from replaying $m$ and getting it accepted by $A$.

If relay $R$ regularly changes keys and discards old private keys, then a replay attack where an adversary first collects keys broadcast by $R$ and then later re-broadcasts these keys in the vicinity of the victim will not work either. By the time the adversary has returned back to the relay $R$, it already discarded the private key necessary to decrypt the replayed messages.

\subsection{Definition and functional requirements}

For two devices $B$ and $C$ define $\contactnow{B}{C}{t}$ to be true when $B$ and $C$ have been within distance $\distance$ of each other for at least $\duration+2\delta$ time at time $t$.\footnote{%
  The additional time $2\delta$ is the leeway we have to allow for the protocols to detect a contact, as it is infeasible in practice to monitor for presence on the Bluetooth channel continuously: this would quickly drain the battery.
}
Define the predicate $\contact{B}{C}$ to be true when $\contactnow{B}{C}{t}$ for some $t$ within the last $\horizon$ days. By definition $\contactnow{B}{C}{t}$ implies $\contactnow{C}{B}{t}$,
and hence $\contact{B}{C}$ implies $\contact{C}{B}$.

\begin{figure*}
  \popetsfigure
  \centering
  \begin{overpic}[abs,unit=1pt]{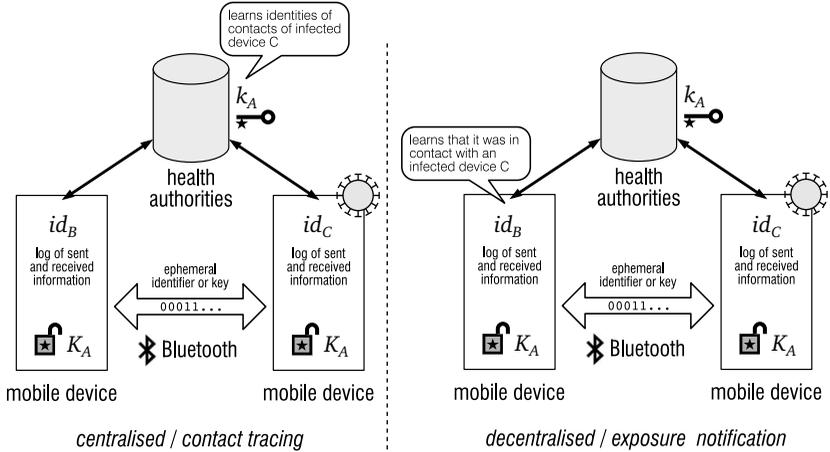}%
    \def\KA{$\pubkey[A]$}
    \def\kA{$\privkey[A]$}
    \def\idB{$\ident[B]$}
    \def\idC{$\ident[C]$}
  \input{./fig/en-ct.overpic}%
  \end{overpic}
  \caption{System model: centralised contact tracing versus decentralised exposure notification.}
  \label{fig-model}
\end{figure*}

Let device $B$ maintains a set
$\contactlist{B} = \Set*{C \mid \contact{B}{C}}$, the set of all devices that $B$ was in contact with (within the last $\horizon$ days).

We can distinguish two different schemes for contact tracing: a \term{centralised} one, and a \term{decentralised} one (see figure~\ref{fig-model}). A centralised scheme for contact tracing allows the health authorities $A$ to obtain $\contactlist{B}$, but only with the active cooperation of a device $B$ itself, and presumably only when the owner of device $B$ tested positive for the virus. 
A decentralised contact tracing system on the other hand allows device $B$ to notify all devices $C$ such that $C \in \contactlist{B}$, but only when $B$ was instructed by the health authorities $A$ to do so.

A decentralised contact tracing system notifies all contacts in $\contactlist{B}$ of an infected person owning device $B$. Such a decentralised contact tracing system is more aptly called a \term{exposure notification} system instead. In a decentralised contact tracing system the health authorities do not learn who has been in contact with an infected person. However, people notified of such an exposure can be instructed to contact the health authorities, so that the health authorities would obtain the same information regardless. In the remainder of this paper we only consider centralised schemes for contact tracing.

We have the following functional requirements for such a contact tracing system~\autocite{dp3t-whitepaper} (where we write `infected device' for the device of a user that tested positive for infection ans subsequently consents to let their device share contact tracing information).
\begin{description}
\item[completeness]
  If $\contact{X}{Y}$ then $Y \in \contactlist{X}$. Whenever device $X$ becomes infected, the health authorities $A$ learn $\ident[X]$ and all $\ident[Y]$ such that $Y \in \contactlist{X}$ (preferably without post-contact cooperation of $Y$).
\item[precision/soundness]
  The health authorities $A$ only learn $\ident[Y]$ when $Y$ itself becomes infected or when $\contact{X}{Y}$ for an infected device $X$ (within the last $\horizon$ days measured from the moment user $X$ tested positive and consented to sharing information). 
\end{description}
In other words, when $Y \notin \contactlist{X}$, the health authorities do not learn $\ident[Y]$. Note that precision of the contact tracing system (\ie the fact that membership of $\contactlist{X}$ corresponds to actual physical proximity for an epidemiologically relevant time period) is pretty much assumed in the remainder of this paper (but please bear in mind the discussion on this important topic in section~\ref{sec-problem}).

\subsection{Threat model}
\label{ssec-threat}

Apart from the devices, their users, and the health authorities we have to reckon with the following additional entities.

Firstly, there may be other devices with access to the (Bluetooth) radio network that are able to eavesdrop on messages exchanged between devices, or that can insert messages of their own. We call these entities \term{adversaries}. We assume that adversaries do not have a valid identity, nor can they fake one from a genuine device.\footnote{%
  Genuine devices may of course also assume this role, but for simplicity we then assume that these are two independent devices.
}
Adversaries are \emph{malicious}.

The contact tracing functionality is implemented by an app running on the devices. We assume the app design and implementation is fully open source and the binary is created using a reproducible build system from these sources. This means we can trust the app to be honest (in so far as we trust the protocols it implements, and that we believe the source is thoroughly scrutinised), but its users could be curious (trying to extract information collected by the app in order to track particular users, or to find out who is infected). In other words devices are \emph{honest but curious}.

Note that actual \term{manufacturers} are responsible for building the devices and providing them with an operating system. These manufacturers are responsible for allowing the contact tracing app to be installed and for the app to access the short range (Bluetooth) radio network (and possible other sensors, data sources and network devices). Although in practice the long range data network (\ie the Internet connection) actually reveals long term network identifiers (like IP addresses) that can be used to track or (re)identify devices, we assume this is not the case here.\footnote{%
  There currently is no way to deal such identifiers in practice if one wants to preserve privacy. A VPN is too weak (the VPN provider sees everything its users do), yet Tor\autocite{DBLP:conf/uss/DingledineMS04} is too strong (there is no need to protect against a NSA like adversary) given the impact on performance. This is yet another example that shows we direly need an efficient, frictionless, way to provide sender anonymity on the Internet, similar to the use of randomised MAC addresses on local networks.
}

\subsection{Security and privacy requirements}
\label{ssec-secpriv}

Given this threat model, we have the following privacy and security requirements~\autocite{dp3t-whitepaper,dp3t-privsec,montjoye2020evaluating-covid19,cho2020contact} for a centralised contact tracing system.
\begin{description}
\item[confidentiality of infection]
  Only the health authorities $A$ learn which devices are infected. In particular, adversaries cannot learn whether (other) devices are infected or not.\footnote{%
    Note that this stronger than claiming that adversaries cannot learn the identity $\ident[Y]$ of infected devices: it also requires them to not be able to single out such a device.
    This is considered an inherent risk of all exposure notification systems as shown in \autocite{dp3t-privsec}, but not of contact tracing systems.
  }
\item[confidentiality of infected contact]
  Suppose $X$ is infected. Apart from the health authorities $A$ no entity learns about a device $Y$ such that $Y \in \contactlist{X}$.
\item[confidentiality of uninfected contacts]
  Suppose $X$ is not infected. No entity learns about a device $Y$ such that $Y \in \contactlist{X}$ (provided $Y$ is not infected). In particular, $X$ does not learn $\ident[Y]$ for any of its `contacts' $Y$.
\item[location privacy]
  Adversaries cannot track a device while it moves around from one physical location to another,\footnote{%
    For example when trying to use the information devices broadcast over the short range broadcast (Bluetooth) radio network in order to detect close by devices and record such contacts for later. It is known that when phones broadcast fixed identifiers, if as few as 1\% of the phones in London would have software to intercept these identifiers, this would allow an attacker to track the location (in a resolution of one square kilometre) of about 54\% of the city’s population~\autocite{montjoye2018cambridge-analytica}.
  }
  either in real-time, or retroactively after additional information is released because of a device becoming infected. This corresponds to systemic risk SR7 identified in~\autocite{dp3t-privsec}. In this paper we show this risk \emph{does not} apply to our protocols (and hence is not really systemic).
\item[authenticity]
  An set of adversaries $\mathcal{M}$ cannot convince a device $X$ that $Y \in \contactlist{X}$ for some device $Y$ for which there is no $t$ such that $\contactnow{X}{Y}{t}$, unless there are $M,M' \in \mathcal{M}$ such that both
  $\contactnow{X}{M}{t}$ and $\contactnow{M'}{Y}{t}$ for some time $t$. In other words, adversaries can only create false positives by relaying information.
\end{description}

\subsection{Other considerations}

There are other (ethical) concerns and considerations that need to be addressed\footnote{%
  See \eg \url{https://rega.kuleuven.be/if/tracing-tools-for-pandemics}
}
when implementing a system for digital contact tracing~\autocite{morley2020ethics-covid19}. These are related to accessibility to the technology used for contact tracing (not everybody has a suitable modern smartphone), whether use of such technology is mandatory, the consequences of having been in close contact of a person that tested positive, or simply the risk of using Bluetooth signal strength as a rather poor proxy for being in close contact~\autocite{dehaye2020bluetooth}. There will be many false positives. These will create `boy-cries-wolf' effects: people being flagged as having been in contact with the virus, but not developing any symptoms, may ignore warnings in the future. Stories about large numbers of such wrongfully flagged people will drive down compliance. Strong enforcement to counter this may backfire.

Some more general requirements for or constraints on contact tracing platforms have been voiced over the past few months. First and foremost is the requirement that the authorities should demonstrate that there is a clear need for the interference with fundamental rights caused by a system like contact tracing, and that this interference is proportionate. This could be done by basing the design on explicit recommendations from the health authority on how to effectively curb the spread of infectious diseases, embedding a system of contact tracing in a larger framework for epidemic control that includes effective testing procedures and quarantine measures. Many more steps can be taken to strengthen the proportionality of the system proposed. For example, a data protection impact assessment should be performed. The system could be developing in consultation with data protection authorities and other stakeholders (including representatives from civil society). Moreover, the designs and implementations should be open. The terms of service should be easily understandable, and there should be a continuous and public mediation/explanation of what is processed, and how. There should also be clear controls for people to exercise their data protection rights, \eg the right to delete data once the epidemic is over, or to see who your data has been shared with.



\section{A peer to peer protocol}
\label{sec-prot1}

We are now ready to present the two centralised contact tracing protocols. The first protocol (discussed in this section) exchanges messages between devices only to create records of proximity. It has one major drawbacks: it involves a handshake between both devices (which creates the risk that a failed handshake causes the proximity of both devices to not be registered). This is resolved by the second protocol, discussed in section~\ref{sec-prot2}.



Each phone $C$ maintains a local log $\logdb[C]{d}$ containing information about other phones detected in the vicinity $d$ days ago, for $d \in \Set{0}{\horizon}$, where $\horizon$ is the maximum number of days for which keeping this data is relevant.


At the start of each day, $\logdb[C]{\horizon}$ is discarded, $\logdb[C]{d} \assign \logdb[C]{d-1}$ for $d = \horizon$ down to $1$, and $\logdb[C]{0} \assign \emptyset$. In other words the log rotates and is pruned daily to remove old entries that are no longer relevant.

\begin{figure}
  \popetsfigure
  \centering
  \begin{overpic}[abs,unit=1pt]{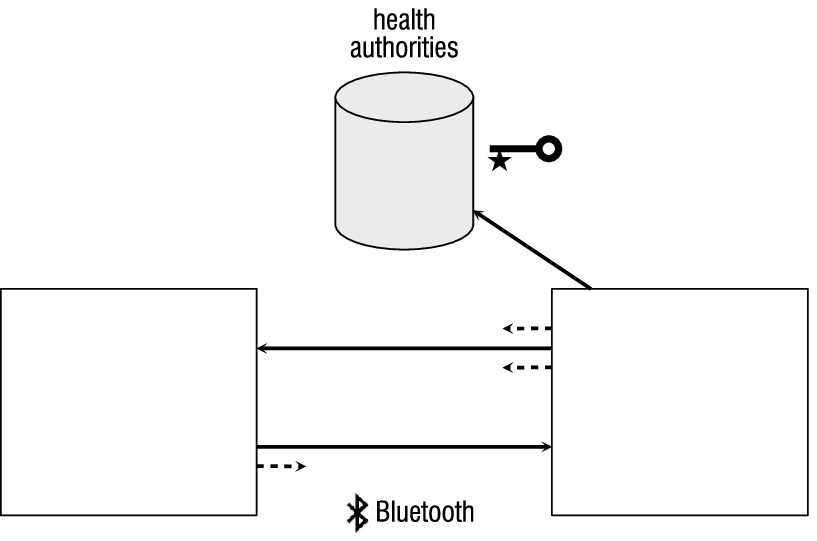}%
    \def\KA{$\pubkey[A]$}
    \def\kA{$\privkey[A]$}
    \def\idB{$\ident[B]$}
    \def\idC{$\ident[C]$}
    \def\KC{$\pubkey[C]$}
    \def\kC{$\privkey[C]$}    
    \def\ECA{$\rrencsym{C}{A}{\ident[B]}{1^\errmargin}$}
    \def\KCECA{\parbox{3cm}{$\pubkey[C]$:\\ $\rrencpart{C}{A}{\ident[B]}$}}
    \def\EA{\KCECA}
  \input{./fig/p2p.overpic}%
  \end{overpic}
  \caption{The peer-to-peer protocol.}
  \label{fig-prot1}
\end{figure}

The full protocol for phones $B$ and $C$ meeting each other now runs as follows (see also figure~\ref{fig-prot1}). The same protocol runs simultaneously with the roles of $B$ and $C$ reversed (and for any other phone that also happens to be in the vicinity). Let $\errmargin$ be a sufficiently large constant, \eg $\errmargin=16$. And let
$\delta$ be some fraction of $\duration$. The protocol broadcasts Bluetooth `beacons' every $\delta$ seconds.
\begin{itemize}
\item
  Whenever its MAC address changes,\footnote{%
     Note that this is not necessarily easy to do in practice.
  }
  $C$ creates a fresh private key $\privkey[C]$ and corresponding public key $\pubkey[C]$. The previous public key is destroyed immediately; the previous private key after $\duration+\delta$ seconds (see below).
\item
  $C$ broadcasts this public key over the Bluetooth network as $\broadcast{key}{\pubkey[C]}$ every $\delta$ seconds.
\item
  When $B$ receives a $\broadcast{key}{\pubkey[C]}$ message, it estimates (based on signal strength) whether the message was sent by another device within unsafe distance $\distance$. If this is the case it stores it in a local pool of keys, together with the time $t$ it first received it. As soon as it received this public key, $B$ encrypts its identity $\ident[B]$ against the public key $\pubkey[A]$ (of the authorities)
  and stores this encrypted identity $\rrencpart{C}{A}{\ident[B]}$ along with $\pubkey[C]$ in the pool. 
  Keys in this pool received more than $\duration+\delta$ seconds ago are discarded, together with their associated data.  
\item
  Every $\delta$ seconds, for every public key $\pubkey[C]$ currently still in this pool, $B$ does the following.
  \begin{itemize}
  \item
    $B$ retrieves the previously stored encryption of its identity for this public key and adds some redundancy $1^\errmargin$ that will allow $C$ to detect successful decryption of a message (see below), and encrypts the message using $\pubkey[C]$.
  \item
    $B$ broadcasts 
    $E = \rrencsym{C}{A}{\ident[B]}{1^\errmargin}
    = \rrenc{C}{A}{\ident[B]}{1^\errmargin}$ over the Bluetooth network as $\broadcast{iam}{E}$. 
  \end{itemize}
\item
  For every $\broadcast{iam}{E}$ messages it receives $C$, $C$ first estimates (based on signal strength) whether the message was sent by another device within unsafe distance $\distance$. If that is the case it tries to decrypt it using its current set of  private keys $\privkey[C]$ (the ones that it generated at most $\duration+\delta$ ago), discarding any results that do not end with $1^\errmargin$.
\item
  If successful, it stores the result $\rrencpart{C}{A}{\ident[B]}$ in its log $\logdb[C]{0}$ for today (together with a copy of $\pubkey[C]$),
  but only if it received at least another copy of the same message at least $\duration$ seconds ago.\footnote{%
     Observe how we here use the fact that $B$ encrypts its identity $\ident[B]$ against the public key $\pubkey[A]$ (of the authorities) \emph{once} for every $\pubkey[C]$ it receives, and stores this encrypted identity $\rrencpart{C}{A}{\ident[B]}$ along with $\pubkey[C]$ in the pool. 
  }
  Note that the log is a set, so duplicates are removed.
\end{itemize}

When someone is diagnosed as being infected, their identity $\ident[C]$ as well as the log $\logdb[C]{}$ is extracted from their phone $C$. The system can be designed such that this requires physical presence of the phone at the site testing for infection, making sure the data is released only after an appropriately authenticated request. This request should come from the medical authorities $A$ or from some independent oversight board. The log $\logdb[C]{}$ is sent to the authorities, who can decrypt the entries
$\rrencpart{C}{A}{\ident[B]}$
using the private key $\privkey[A]$ to obtain the set of all identities $\ident[B_i]$ that have been in the vicinity of $C$ in the last $\horizon$ days. $A$ verifies that the hash of $\pubkey[C]$ is indeed present to thwart replay attacks.

$B$ relies on $C$ to discard old log entries, and to only record its encrypted identity if it was at an unsafe distance for more than $\duration$ seconds.

After releasing its log, the app locks itself: it no longer records any data. Reinstalling the app to reactivate it is (made) impossible. We are assuming here that once a person has been shown to be infected, he or she will stay immune and therefore not spread the virus further. Thus tracing who has been close to this person is no longer necessary.

\subsection{Analysis}

Define $\contactlist{X}$ as the set of all $Y$ such that
$\rrencpart{X}{A}{\ident[Y]}$ in $\logdb[X]{}$. We analyse the requirements outlined in section~\ref{sec-problem} one by one.

\paragraph{Completeness}

We have to show that if $\contact{X}{Y}$ then $Y \in \contactlist{X}$, and that
whenever device $X$ becomes infected, the health authorities $A$ learn $\ident[X]$ and all $\ident[Y]$ such that $Y \in \contactlist{X}$.

The first part is shown as follows. Suppose $\contact{X}{Y}$. This means that
$X$ and $Y$ have been within distance $\distance$ of each other for at least $\duration+2\delta$ time at some time $t$. $X$ broadcasts $\pubkey[X]$ every $\delta$ seconds, so $Y$ has received $\pubkey[X]$ at time $t-\duration-\delta$ at the latest. This means that during the time interval $[t-\duration-\delta,t]$ the pool of $Y$ contains $\pubkey[X]$
and $X$ has kept a copy of the corresponding private key $\privkey[X]$. Every $\delta$ seconds within this time interval, $Y$ broadcasts $E = \rrencsym{X}{A}{\ident[Y]}{1^\errmargin} = \rrenc{X}{A}{\ident[Y]}{1^\errmargin}$
which $X$ is assumed to receive as it is within distance $\distance$ during this interval. $X$ can decrypt this message as it kept a copy of the corresponding private key $\privkey[X]$. The first and last such message $X$ receives during this interval are guaranteed to be at least $\duration$ time apart, and are guaranteed to be equal as $Y$ encrypts its identity $\ident[Y]$ against the public key $\pubkey[A]$ (of the authorities) \emph{once} for every $\pubkey[X]$ it receives. This guarantees that $X$ includes $\rrencpart{C}{A}{\ident[Y]}$ in $\logdb[X]{}$ as required. 

The second part follows from the fact that once $X$ is diagnosed as infected, its log $\logdb[X]{}$ is uploaded to the authorities. As this contains all entries
$\rrencpart{X}{A}{\ident[Y]}$ such that $Y \in \contactlist{X}$, and the authorities can decrypt this using their knowledge of $\privkey[A]$, the result follows.

\paragraph{Precision/soundness}

We have to show that the health authorities $A$ only learn $\ident[Y]$ when $Y$ itself becomes infected or when $\contact{X}{Y}$ for an infected device $X$ (within the last $\horizon$ days measured from the moment user $X$ tested positive and consented to sharing information).

First observe that according to the protocol, only infected devices $X$ send their identifier (and their logs) to the authorities directly. Moreover, a device $Y$ only sends its identifier to other devices whose key $\pubkey[X]$ was recently received (these keys are discarded after $\duration+\delta$ seconds) while within the unsafe distance $\distance$. The identity $\ident[Y]$ is sent encrypted, first against the key of the authorities, and then again using this
key $\pubkey[X]$. Any device that has not been within this unsafe distance of $Y$ will therefore not be able to decrypt this message.

Note however that a device $X$ will already add $\xenc{A}{\ident[Y]}$ in its log $\logdb[X]{0}$ for today if it receives two copies of the same message at least $\duration$ seconds apart. This is a strong indication that $X$ and $Y$ were within unsafe distance $\distance$ during that whole period, but does not guarantee that during that period $X$ and $Y$ were actually that close all the time. Precision therefore only holds approximately.

\paragraph{Confidentiality of infection}

We have to show that only the health authorities $A$ learn which devices are infected. This immediately follows from the fact only when someone is diagnosed as being infected, their identity $\ident[X]$ as well as the log $\logdb[X]{}$ is extracted from their phone $X$ and sent to the authorities, and that no information (about infection status) leaves the authorities.

Observe that the authorities may learn how often device $Y$ was in close contact with device $X$ during the day (or, at the very least, on how many different days).

\paragraph{Confidentiality of infected contact}

Suppose $X$ is infected. We have to show that, apart from the health authorities $A$, no entity learns about a device $Y$ such that $Y \in \contactlist{X}$.

We defined $\contactlist{X}$ as the set of all $Y$ such that
$\rrencpart{X}{A}{\ident[Y]}$ in $\logdb[X]{}$. As observed above, $\logdb[X]{}$ is only shared to the authorities and clearly $X$ itself knows $\logdb[X]{}$.

If we can show that even $X$ learns nothing about a device $Y$ such that $Y \in \contactlist{X}$, then no other entity can either.

Note that all $X$ receives are
$\rrencsym{X}{A}{\ident[Y]}{1^\errmargin} = \rrenc{X}{A}{\ident[Y]}{1^\errmargin}$
messages which it can decrypt only if the correct key $\pubkey[X]$ is used. This yields $\rrencpart{X}{A}{\ident[Y]}$.  

\paragraph{Confidentiality of uninfected contacts}

Suppose $X$ is not infected. We have to show that no entity learns about a device $Y$ such that $Y \in \contactlist{X}$. This follows from the analysis in the previous paragraph.

\paragraph{Location privacy}

We have to show that an adversary cannot track a device while it moves around from one physical location to another.

A device $X$ either broadcasts a random key $\pubkey[X]$, or an encrypted identifier $\rrenc{Y}{A}{\ident[X]}{1^\errmargin}$.

Regarding the key, the protocol guarantees/assumes it is refreshed whenever the MAC address of the Bluetooth channel changes. In other words, broadcasting this key does not create an additional vulnerability on top of the location tracking issues associated with MAC address randomisation~\autocite{martin2017mac-addr-randomisation}

Regarding the encrypted identifier, note that for any entity not equal to $Y$ (to whose key the identifier is encrypted) the message is essentially random 
as the ElGamal encryption used is semantically secure (see section~\ref{sec-preliminaries}). Device $Y$ can decrypt (and any device in $X$ vicinity can send its own random key and receive and decrypt) this message to obtain $\xenc{A}{\ident[X]}$. Depending on the cipher used for this encryption (semantically secure or not), $Y$ can observe a certain device to be in its vicinity during the day or not (see the discussion earlier on confidentiality of infected contact).

\paragraph{Authenticity}

We have to show that a set of adversaries $\mathcal{M}$ cannot convince a device $X$ that $Y \in \contactlist{X}$ for some device $Y$ for which there is no $t$ such that $\contactnow{X}{Y}{t}$, unless there are $M,M' \in \mathcal{M}$ such that both $\contactnow{X}{M}{t}$ and $\contactnow{M'}{Y}{t}$ for some time $t$.

This trivially follows from the fact that for any device $Y$, the value $\ident[Y]$ is secret (it is only revealed to the authorities) and cannot be guessed. The authorities never reveal any information they receive. Devices only broadcast their identity using replay resistant encryption. This ensures that no other entity ever obtains another device identity in plaintext, and that the authorities do not accept a replayed encryption. Therefore old messages that are replayed will be ignored, and ensures that we are in a relay scenario where $M,M' \in \mathcal{M}$ such that both $\contactnow{X}{M}{t}$ and $\contactnow{M'}{Y}{t}$ for some time $t$.

\section{Involving the help of a central server}
\label{sec-prot2}

The peer-to-peer protocol presented in the previous section has a significant drawback: the protocol involves a handshake where phone $C$ broadcasts a random public key and phone $B$ responds with some encrypted information. If either of these messages fail to arrive, a possible contact remains undetected. In other words, the protocol is error-prone. 

The following protocol solves these issues by involving the help of a central server. This help comes at a price though: more care has to be taken to ensure that the authorities do not learn more than necessary.

Let $\hash : \hashdom \mapsto \hashdom$ be a cryptographic hash function from a domain $\hashdom$ to a range $\hashdom$.

As before each phone $C$ maintains a local log $\logdb[C]{d}$ containing information about the keys it used $d$ days ago, for $d \in \Set{0}{\horizon}$. This log rotates and is pruned daily to remove old entries that are no longer relevant, as described for the previous protocol.

To reliably detect close proximity for longer than the safe period $\duration$, time is divided into epochs $\epoch{i}$ that start at time $i \cdot \duration$ and last for $\duration$ seconds.\footnote{%
  Some form of synchronisation of clocks between devices is needed to prevent replay attacks. The closer the synchronisation, the harder replay becomes. See the discussion later on.
}
If two device are found to have been in close proximity in two consecutive epochs, this is assumed to be a strong indicator that there is a significant risk of infection. This less exact method of establishing the duration of a particular contact is used to prevent the central server of the authorities from determining when exactly this contact took place.

\begin{figure*}
  \popetsfigure
  \centering
  \begin{overpic}[abs,unit=1pt]{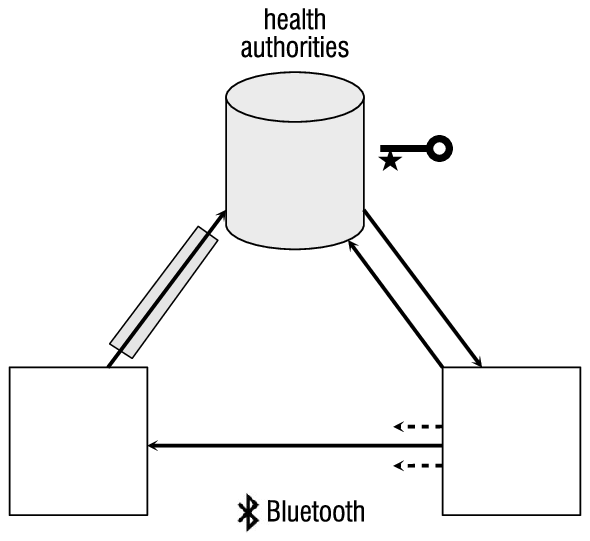}%
    \def\KA{$\pubkey[A]$}
    \def\kA{$\privkey[A]$}
    \def\idB{$\ident[B]$}
    \def\idC{$\ident[C]$}
    \def\KC{$\pubkey[C],\sign{C}{t}$}
    \def\kC{$\privkey[C]$}    
    \def\ei{$\epochrand{i}$}
    \def\KCECA{\parbox{4cm}{$\pubkey[C]$:\\ $\xenc{C}{\xenc{A}{\ident[B]}  \| \epochrand{i} \| \hash(\epochrand{i-1})}$}}
    \def\ECA{$\xenc{C}{\xenc{A}{\ident[B]} \| \epochrand{i} \| \hash(\epochrand{i-1})}$}
    \def\EA{$\xenc{A}{\ident[B]}$}
  \input{./fig/cs.overpic}%
  \end{overpic}
  \caption{The protocol involving the central server.}
  \label{fig-prot2}
\end{figure*}

The full protocol for phones $B$ and $C$ meeting each other now runs as follows (see also figure~\ref{fig-prot2}). 
\begin{itemize}
\item
  $B$ generates a random value $\epochrand{i} \rassign \hashdom$ for each epoch $\epoch{i}$; it keeps this random for one more epoch and then discards it. 
\item
  Whenever its MAC address changes, $C$ creates a fresh private key $\privkey[C]$ and corresponding public key $\pubkey[C]$. It stores these keys in the local log $\logdb[C]{0}$ for today.
\item
  $C$ regularly (at least every $\delta$ seconds) broadcasts $\broadcast{key}{\pubkey[C]}$ over the Bluetooth network.
  If replay must be detected, it also sends the current time, signed with the current private key $\sign{C}{t}$. The drawback is that in this case the broadcast no longer fits in a single Bluetooth beacon.
\item
  $B$, when receiving such a broadcast, first estimates (based on signal strength) whether the message was sent by another device within unsafe distance.
\item
  If this is the case, and the message contains a signature $\sign{C}{t}$, $B$ verifies this signature with the public key $\pubkey[C]$ just received, and if the verification is successful, checks that the time $t$ is equal to its own estimate of the current time.  
\item
  If this is the case, $B$ encrypts its identity $\ident[B]$ using the public key $\pubkey[A]$ (of the authorities), it adds the current epoch random $\epochrand{i}$ and the hash $\hash(\epochrand{i-1})$ of the previous epoch random first and then encrypts the result using the public key $\pubkey[C]$ it just received.
\item
  $B$ sends the result
  $\xenc{C}{\xenc{A}{\ident[B]} \| \epochrand{i} \| \hash(\epochrand{i-1})}$ together with the public key $\pubkey[C]$ to the authorities $A$ using the cellular network (again encrypted against the public key of $A$, not further shown here). Note that we assume here that $B$'s identity cannot be recovered by inspecting the cellular network, see section~\ref{ssec-threat}.
\item
  The authorities $A$ store this information in their database, indexed using the public key $\pubkey[C]$ used to encrypt it. Information older than $\horizon$ days should automatically be pruned (as this information can never be decrypted anymore, see below).
\end{itemize}

In this protocol, the log $\logdb[C]{}$  contains all private/public key pairs $\privkey[C],\pubkey[C]$ that $C$ broadcast and that have been used by all phones in the vicinity of $C$ the last $\horizon$ days to encrypt their identities $\ident[B_i]$ before submitting them to the authorities. When someone is diagnosed as being infected, \emph{only the public keys} in the log are extracted from their phone $C$ (similar to the previous protocol), and sent to the authorities.  This allows the authorities to search for any entries
$\xenc{C}{\xenc{A}{\ident[B]} \| \epochrand{i} \| \hash(\epochrand{i-1})}$
stored in their database indexed by the public key $\pubkey[C]$. All entries found are sent to device $C$ which uses the corresponding private keys $\privkey[C]$ stored in the log to decrypt.

As a result, device $C$ obtains entries
$\xenc{A}{\ident[B]} \| \epochrand{i} \| \hash(\epochrand{i-1})$
for all devices $B$ that have been in close contact with $C$.
To determine whether this contact was sufficiently long, $C$ looks for any two entries $\xenc{A}{\ident[B]} \| \epochrand{i} \| \hash_i$ and
$\xenc{A}{\ident[B]} \| \epochrand{j} \| \hash_j$ (collected for the private and public keys used on the same day $d$) such that $\hash_i = \hash(\epochrand{j})$ (or vice versa of course). The precision of this approach can be improved if one makes each epoch shorter, and requires the matching process to find a sufficiently long chain of epochs that together cover at least $\duration$ seconds (the minimum time before a contact is deemed sufficiently risky). For any such entries found it returns $\xenc{A}{\ident[B]}$ to the authorities. Using their knowledge of $\privkey[A]$ they can decrypt this to recover $\ident[B]$.

$B$ itself ensures that it only reports its identity if it sees another device $C$ to be closer than the unsafe distance. $B$ relies on $C$ to only return 
$\xenc{A}{\ident[B]}$ when this encrypted identity is found to be linked to consecutive epochs, and not to leak information about the time and the actual length of the encounter (which $C$ could deduce based on its knowledge of when it used a particular key, and using repeated chaining of epoch hashes).

\subsection{Analysis}

Let DB be the database of all entries
$\xenc{X}{\xenc{A}{\ident[Y]} \| \epochrand{i} \| \hash(\epochrand{i-1})}$
stored by the authorities. In this protocol $\contactlist{X}$ is defined to be all $Y$ such that DB contains two entries
$\xenc{X}{\xenc{A}{\ident[Y]} \| \epochrand{i} \| \hash_i}$ and
$\xenc{X'}{\xenc{A}{\ident[Y]} \| \epochrand{j} \| \hash_j}$
encrypted using two public keys $\pubkey[X]$ and $\pubkey[X']$ used by device $X$ on the same day such that $\hash_i = \hash(\epochrand{j})$ (or vice versa of course). 

We again analyse the requirements outlined in section~\ref{sec-problem} one by one.

\paragraph{Completeness}

We have to show that if $\contact{X}{Y}$ then $Y \in \contactlist{X}$, and that
whenever device $X$ becomes infected, the health authorities $A$ learn $\ident[X]$ and all $\ident[Y]$ such that $Y \in \contactlist{X}$.

The first part is shown as follows. Suppose $\contact{X}{Y}$. Then there is a time $t$ such that $X$ and $Y$ have been within distance $\distance$ of each other for at least $\duration+2\delta$ time. Time is divided into epochs $\epoch{i}$ that start at time $i \cdot \duration$ and last for $\duration$ seconds. As $X$ broadcasts its current public key $\pubkey[X]$ at least every $\delta$ seconds, this means $Y$ must receive $X$'s current public key in at least two consecutive epochs $\epochrand{i-1}$ and $\epochrand{i}$ with sufficient signal strength. This means $Y$ sends
$\xenc{X}{\xenc{A}{\ident[Y]} \| \epochrand{i-1} \| \hash(\epochrand{i-2})}$ and later
$\xenc{X}{\xenc{A}{\ident[Y]} \| \epochrand{i} \| \hash(\epochrand{i-1})}$
to the authorities. $X$'s key may have changed during this period. Regardless, the conditions for $Y \in \contactlist{X}$ are satisfied.

The second part follows from the fact that once $X$ is diagnosed as infected, 
it sends its public keys $\pubkey[X]$ in its log to the authorities, who use it to look up all entries $\xenc{X}{\xenc{A}{\ident[Y]} \| \epochrand{i} \| \hash(\epochrand{i-1})}$ in DB. These are sent to $X$ for decryption. $X$ can decrypt these as long as they are not older than $\horizon$ days, because that is how long $X$ keeps the private keys $\privkey[X]$ it used in its logs.
We already established that both
$\xenc{X}{\xenc{A}{\ident[Y]} \| \epochrand{i-1} \| \hash(\epochrand{i-2})}$ and 
$\xenc{X}{\xenc{A}{\ident[Y]} \| \epochrand{i} \| \hash(\epochrand{i-1})}$
are in DB. This means $X$ obtains
$\xenc{A}{\ident[Y]} \| \epochrand{i-1} \| \hash$ and
$\xenc{A}{\ident[Y]} \| \epochrand{i} \| \hash$. As $\hash(\epochrand{i-1})=\hash'$ we have a match and $X$ returns $\xenc{A}{\ident[Y]}$ to the authorities, who can now decrypt it to recover $\ident[Y]$ as required. They also learn $\ident[X]$ as soon as $X$ tests positive.

\paragraph{Precision/soundness}

We have to show that the health authorities $A$ only learn $\ident[Y]$ when $Y$ itself becomes infected or when $\contact{X}{Y}$ for an infected device $X$ (within the last $\horizon$ days measured from the moment user $X$ tested positive and consented to sharing information).

First observe that according to the protocol, only infected devices $X$ send their identifier to the authorities directly. Other devices $Y$ only send
$\xenc{X}{\xenc{A}{\ident[Y]} \| \epochrand{i} \| \hash(\epochrand{i-1})}$ whenever they receive a public key from another device $X$ that they detect to be within unsafe distance $\distance$. These messages cannot be decrypted after $\horizon$ days because $X$ purges keys older than that from its log.

As we have seen above, the authorities need the help of an infected device $X$ to decrypt such messages. If not $\contact{X}{Y}$, it is unlikely (though not impossible if $X$ and $Y$ are close around an epoch rollover) that $Y$ sent two messages
$\xenc{X}{\xenc{A}{\ident[Y]} \| \epochrand{i-1} \| \hash(\epochrand{i-2})}$ and
$\xenc{X}{\xenc{A}{\ident[Y]} \| \epochrand{i} \| \hash(\epochrand{i-1})}$
for two consecutive epochs, which would trigger $X$'s rule to return
$\xenc{A}{\ident[Y]}$ to the authorities. 
Again precision only holds approximately.

\paragraph{Confidentiality of infection}

We have to show that only the health authorities $A$ learn which devices are infected.

This immediately follows from the fact only when someone is diagnosed as being infected, their identity $\ident[X]$ is extracted from their phone $X$ and sent to the authorities, and that no information (about infection status) leaves the authorities. Throughout the protocol the identity of a device is always encrypted against the key of the authorities.

Note that the authorities \emph{do} learn how often (over the full period of $\horizon$ days) an infected device $X$ was in close contact with another device $Y$ (but not on which exact time or on which days in particular).

\paragraph{Confidentiality of infected contact}

Suppose $X$ is infected. We have to show that, apart from the health authorities $A$, no entity learns about a device $Y$ such that $Y \in \contactlist{X}$.

Note that $X$ (only) receives entries
$\xenc{X}{\xenc{A}{\ident[Y]} \| \epochrand{i} \| \hash(\epochrand{i-1})}$
which it can decrypt to obtain
$\xenc{A}{\ident[Y]} \| \epochrand{i} \| \hash(\epochrand{i-1})$. As
$\xenc{A}{\ident[Y]}$ is semantically secure, this does not leak information about $\ident[Y]$ (or the frequency with which $\ident[Y]$ occurs in the entries that $X$ receives).
Again, throughout the protocol the identity of a device is always encrypted against the key of the authorities.

\paragraph{Confidentiality of uninfected contacts}

Suppose $X$ is not infected. We have to show that no entity learns about a device $Y$ such that $Y \in \contactlist{X}$. 

This trivially follows from the fact that in that case $X$ is not asked to match any data on behalf of the authorities. Then the authorities store
$\xenc{X}{\xenc{A}{\ident[Y]} \| \epochrand{i} \| \hash(\epochrand{i-1})}$
safely encrypted against both $\pubkey[A]$ and $\pubkey[C]$. 
Again, throughout the protocol the identity of a device is always encrypted against the key of the authorities.

\paragraph{Location privacy}

We have to show that an adversary cannot track a device while it moves around from one physical location to another.

A device $X$ either broadcasts a random key $\pubkey[X]$, a signed timestamp $\sign{X}{t}$ 
or an encrypted identifier
$\xenc{X}{\xenc{A}{\ident[Y]} \| \epochrand{i} \| \hash(\epochrand{i-1})}$

For the random public key, we again assume it is refreshed whenever the MAC address of the Bluetooth channel changes. The encrypted identifier does not pose an issue as the message is encrypted using a semantically secure cipher. In both cases, a new message can never be linked to the previous message from the same device. The signature is always different because the time changes. When clocks are perfectly synchronised, and change MAC address at the very same time, two different devices both send the same signature. If however, clocks are not perfectly synchronised and MAC addresses do not change at exactly the same time, then two devices could be traced because one device emits a sequence
$\sign{X}{t},\sign{X}{t+\omega},,\sign{X}{t+2\omega},\ldots$ while the other
emits a sequence
$\sign{X}{t'},\sign{X}{t'+\omega},,\sign{X}{t'+2\omega},\ldots$ 

\paragraph{Authenticity}

We have to show that a set of adversaries $\mathcal{M}$ cannot convince a device $X$ that $Y \in \contactlist{X}$ for some device $Y$ for which there is no $t$ such that $\contactnow{X}{Y}{t}$, unless there are $M,M' \in \mathcal{M}$ such that both $\contactnow{X}{M}{t}$ and $\contactnow{M'}{Y}{t}$ for some time $t$.

This follows from the fact that $Y$ never sends a message to the authorities encrypted against a public key $\pubkey[X]$ that wasn't recently (according to the signature $\sign{X}{t}$) sent by $X$ itself. This means that either $X$ is close to $Y$, or its messages are relayed in real time by some adversarial devices.

\section{Related work}
\label{sec-related}

Many papers have been published last year proposing some protocol for contact tracing or exposure notification. Martin~\etal\autocite{martin2020demystifying-covid19-tracing} offer an excellent overview of the state of the art. We highlight the main Bluetooth-based approaches here and compare them with our proposal.

One of the first protocols of these type was the TraceTogether app deployed in Singapore~\footnote{%
  \url{https://www.tracetogether.gov.sg}
}.
This inherently centralised approach lets phones exchange regularly changing pseudonyms over Bluetooth. A phone of a person who tests positive is requested to submit all pseudonyms it collected to the central server of the health authorities, who are able to recover phone numbers and identities from these pseudonyms. This allows them to quickly contact anybody in close contact with this infected person. Privacy is clearly less of a concern. \footnote{%
  The Singapore authorities recently announced that the police can access COVID-19 contact tracing data for criminal investigations. \url{https://www.zdnet.com/article/singapore-police-can-access-covid-19-contact-tracing-data-for-criminal-investigations/}
}

Other notable early protocols are Cho~\etal\autocite{cho2020contact} (that use an anonymous message passing approach to allow devices to send infection status messages to other devices that were in close proximity), and
Canetti~\etal\autocite{canetti2020anonymous} that is very similar to the DP-3T to be discussed next.

\subsection{The DP-3T protocol}

The Decentralized Privacy-Preserving Proximity Tracing (DP-3T) protocol\footnote{%
  \url{https://github.com/DP-3T/documents}
}
is a decentralised protocol for exposure notification~\autocite{dp3t-whitepaper}. The protocol uses locally generated frequently changing ephemeral identifiers (EphIDs) that devices broadcast via Bluetooth Low Energy advertisements. Other devices store the EphIDs they observe, together with the duration and a coarse indication of the time of contact. In the so called ``Low-Cost'' design of DP-3T, the device of an infected user uploads all EphIDs \emph{it itself generated} to a central server. Other devices regularly query this server to download any new EPhIDs and locally match these with EphIDs received earlier from devices in close proximity. Any match indicates a contact, of which the user is notified by the app itself. The central server does not learn these matches. A variant of this protocol uses Cuckoo hashing
to hide the actual EphIDs of infected users, offering only the resulting Cuckoo filter for download to other devices that want to check whether they have been in close contact with an infected user. This is done to somewhat mitigate the otherwise existing risk that the EphIDs of infected users revealed can be used to retroactively track their location~\autocite{vaudenay2020dp3t}.

The DP-3T consortium has done a tremendous amount of work in an incredible short amount of time by creating specifications~\autocite{dp3t-whitepaper}, reference implementations\footnote{%
  \url{https://github.com/DP-3T/}
},
and a detailed risk analysis~\autocite{dp3t-privsec}. Their efforts, also influencing the policy agenda of the European Commission and beyond, are a shining example for all of us working in the area of privacy enhancing technologies. The short description given here barely does their work justice.

Google and Apple released an interoperable framework for exposure notification (GAEN) based on the DP3T protocols. Elsewhere we argue that this creates a dormant functionality for mass surveillance at the operating system layer, and  show how it does \emph{not} technically prevent the health authorities from implementing a purely centralised form of contact tracing (even though that is the stated aim)~\autocite{hoepman2020gaen-critique}.

The main advantage of the protocols discussed here (over DP3T and GAEN) is that information revealed by infected users cannot be used to retroactively track their location. Moreover, our protocols solve contact tracing, not exposure notification.

\subsection{The DESIRE protocol}
\label{ssec-desire}

The DESIRE protocol\footnote{%
  \url{https://github.com/3rd-ways-for-EU-exposure-notification/project-DESIRE}
}
(designed by INRIA PRIVATICS TEAM as a a follow-up to their ROBERT protocol\footnote{%
  \url{https://github.com/ROBERT-proximity-tracing}
})
is a hybrid solution for contact tracing where determining the risk of infection happens centrally. Instead of collecting (and sharing) temporary pseudonyms (as \eg the DP3T protocol does), DESIRE is based on Private Encounter Tokens (PETs) instead, that remove the tracking risks inherent to decentralised solutions based on pseudonyms, and that also mitigates against replay attacks. These PETs are essentially Diffie-Hellman based shared secrets generated as follows. Devices $A$ and $B$ regularly generate a new device secret $s_A$ and $s_B$ and use that to generate Ephemeral Bluetooth Identifiers (EBID) $E_A = g^{s_A}$ and $E_B = g^{s_B}$ (within some suitable group with generator $g$). Devices broadcast these EBIDs using Bluetooth. When receiving an EBID $E_A$, device $B$ computes a Private Encounter Tokens $P_{BA} = H(E_A^{s_B}) = H(g^{s_As_B})$. Device $A$ similarly computes $P_{AB}=P_{BA}$ when receiving $E_B$. Devices of infected users upload all PETs they collected to the central server. Devices of other users regularly upload a list of collected PETs they collected to allow the central server to compute a risk score.

This means that DESIRE is different from our protocols in that it requires the active participation of devices that have been in contact of an infected device not only during the moment of contact itself, but also \emph{post-contact} after the infected device was notified of being infected. DESIRE also assumes synchronised clocks (which our first protocol does not rely on).

DESIRE also requires a handshake (\ie a successful bidirectional exchange of messages between two proximate devices) for a contact to be successfully registered. Our second protocol thus also improves on the DESIRE protocol in this respect. Our protocols allow similar techniques to be deployed as described in the DESIRE proposal to prevent the authorities to learn which log entries belong to which particular infected user (thus hiding even the social graph of infected users from the authorities). For example, log entries from several devices could be uploaded simultaneously through a mixing network~\autocite{chaum1981untraceable-mail}.

\section{Conclusion}
\label{sec-conclusion}

We have proposed two centralised protocols for digital contact tracing that, contrary to the common hypothesis that this is an inherent risk, do not allow (retroactive) tracking of the location of a device over time. We have done so even though we have strong reservations against the use of contact tracing in fighting the epidemic (given the significant privacy infringements it causes and the limited effectiveness to expect of it), as we expect governments to implement such systems for contact tracing anyway. In that case, it is important to have relatively comprehensive map of the design space available to guide them in their choice.

We thank the anonymous referees for their very insightful comments that helped to improve the protocols and the presentation.

%
%
%
\setlength{\emergencystretch}{8em}
\printbibliography
\end{document}